\documentclass[preprint,12pt,sort&compress]{elsarticle}

\usepackage{epsfig}

\usepackage{amssymb}
\usepackage{amsthm}

\usepackage{lineno}

\journal{Optics  communications}

\begin{document}

\begin{frontmatter}



\title{Conditional phase gate and quantum state transfer
via off-resonant quantum Zeno dynamics}


\author{Wan-Jun Su}
\ead{wanjunsu@fzu.edu.cn}
\address{Department of Physics, Fuzhou University, Fuzhou
350002, People's Republic of China}

\begin{abstract}
  We propose a scheme to realize the conditional phase gate (CPG) and quantum state transfer (QST) between two qubits (acted by nitrogen-vacancy (NV) centers) based on off-resonant quantum Zeno dynamics. We also consider the entanglement dynamics of two qubits in this system. Since no cavity photons or excited levels of the NV center is populated during the whole process, the scheme is immune to the decay of cavity and spontaneous emission of the NV center. The strictly numerical simulation shows that the fidelities of QST and CPG are high even in the presence of realistic imperfections.
\end{abstract}

\begin{keyword}
\ { phase gate,quantum Zeno effect,N-V center}
\PACS {03.65.Xp,03.65.Vf,42.50.Dv}
\end{keyword}

\end{frontmatter}



\section{Introduction }
\label{}
Reliable quantum state transfer (QST) and quantum phase gates between
distant qubits have become a holy grail of quantum physics research,
owing to its potential application in a scalable quantum information
processing (QIP) \cite{Lloyd}. So far, many proposals have been
presented for QIP in different quantum systems \cite{Zheng,Bose,Moehring,Kielpinski,Blinov,Greentree,yang,You}.
Among these systems, the composite systems of nitrogen-vacancy (NV)
centers embedded in nanocavities are considered to be a fresh tool
for a room-temperature solid-state QIP \cite{Bernien}. The NV center
in diamond has a long electron spin coherence time and can be manipulated by the optical or microwave pulses \cite{Epstein}. Since the ability to address individual NV center
was proved, a lot of theoretical and experimental efforts have been
devoted to QIP based on the composite microcavity-NV center
systems \cite{Schietinger,Kimble,Childress,Sean}. The significant
advances in implementing various protocols for QIP can lead to
long-distance quantum communication or the creation of a quantum
computer in the future. Especially, Li et al \cite{Li} have proposed a scheme for the realization of QST and entanglement with NV centers coupled to a high-Q whispering gallery
modes (WGM) microresonator, which is processed via Raman transitions. Based on the weak coupling limit for pure state transport, Ajoy and Cappellaro \cite{Ajoy} have proposed a scheme for realizing perfect quantum transport in two separated NV centers.

The quantum Zeno effect \cite{Misra,Cook,Itano} occurs when a slowly
evolving quantum system undergoes a rapid sequence measurement.
In 2002, Facchi and Pascazio found the quantum Zeno dynamics
\cite{Facchi}, which took continuous couplings as a substitute for frequent measurements and hindered the evolution of the quantum system.
Until now, the new finding has enlightened numerous schemes
\cite{A,R,Shen} to implement quantum gates, prepare quantum
entangled states and transfer quantum information. For example, Shi
et al \cite{shi} have shown how to realize quantum information transfer for one atom to
another in coupled-cavity system via quantum Zeno dynamics. Based on
the quantum Zeno dynamics, these protocols above are robust against
cavity decay. However, atomic spontaneous emission may affect these systems.

For the purpose of preventing the decoherence including the atomic
spontaneous emission and cavity decay, off-resonant quantum Zeno dynamics should be
used to remedy the defect in the resonant case \cite{F}. For instance, Zhang et al \cite{shao} proposed a scheme to realize a $\sqrt{swap}$ gate via off-resonant quantum Zeno dynamics. Inspired by their works above, we will propose a scheme to implement QST and
CPG between NV centers via the off-resonant quantum Zeno dynamics.
Moreover, the entanglement dynamics of two qubit off-resonantly
coupled to a cavity with weak driving classical fields will also be
considered in the present paper. The present approach has the
following merits: (i) The cavity field would not be excited in the whole
process, and the interaction is a virtual-photon process. In other
words, our model works well in the bad-cavity limit, which makes it
more applicable to current laboratory techniques. (ii) The idea,
which combines the advantages of the quantum Zeno dynamics and the
large detuning between NV centers and cavity, will make the
protocols also robust against NV center spontaneous decay.

\section{model and effective dynamics}
\label{}

We begin by considering a system composed of two separated NV centers
simultaneously interacting with a microcavity (e.g., microtoroidal
resonator). The microcavity with high quality factor, serves as a
quantum bus, as shown in Fig.1. The states of the NV center form a
$\wedge$ configuration with the ground states
$|m_s=+1\rangle=|g\rangle$, $|m_s=-1\rangle=|f\rangle$ and the
excited state $\left| A_2\right\rangle=(\left|
E_{-}\right\rangle \left| +1\right\rangle +\left|
E_{+}\right\rangle \left| -1\right\rangle )/{\sqrt{2}}=\left| e\right\rangle$,
where $\left| E_{\pm }\right\rangle $ are orbital states with angular
momentum projection $\pm1$ along the N-V axis. Another ground state
$|m_s=0\rangle=|i\rangle$ is an ancillary state. The transition $
\left|f \right\rangle\rightarrow \left|e \right\rangle$ is driven by
a microwave field with Rabi frequency $\Omega$ and detuning
$\Delta_{d}$. While, the transition $ \left|g
\right\rangle\rightarrow \left|e \right\rangle$ couples to the
cavity field with coupling strength $g_i$ (where $i=1,2$), and the
corresponding detuning is $\Delta_{c}$. For convenience, we assume $g_1
= g_2 = g$ and $\Delta_d = \Delta_c = \Delta$ to be real. In the
frame rotating with the cavity frequency $\omega_c $, the
Hamiltonian of the combined system is given by ($\hbar=1$),

\begin{eqnarray}\label{eq1-1}
H&=&H_{c}+H_{l}+H_{de},\nonumber\\
H_{c}&=&\sum_{i=1,2} g_{i}(|e\rangle_i\langle g|a+|g\rangle_i\langle e|a^{+}),\nonumber\cr
H_{l}&=&\sum_{i=1,2}\Omega_i(|e\rangle_i\langle f|+|f\rangle_i\langle e|),\nonumber\cr
H_{de}&=&\Delta(|e\rangle_1\langle e|+|e\rangle_2\langle e|),
\end{eqnarray}

where $a^{+}$ and $a$ are the creation and annihilation operators
for the cavity mode, respectively. $H_{c}$ stands for the NV
center-cavity interaction, and $H_{l}$ represents the interaction
between the NV centers and the classical field.

If the initial state of the system is $|g f\rangle|0\rangle_c$, it
will evolve in a close subspace spanned by $\{|g
f\rangle|0\rangle_c, |g e\rangle|0\rangle_c, |gg\rangle|1\rangle_c,
|e g\rangle|0\rangle_c, |fg\rangle|0\rangle_c\}$. Therefore the
above Hamiltonian can be rewritten with eigenstates of the $H_c$
representation:
\begin{eqnarray}\label{1}
H&=&H_{c}+H_{l}+H_{de}\nonumber\cr
H_{c}&=&-\sqrt{2}g(|\psi_2\rangle\langle \psi_2|-
|\psi_3\rangle\langle
 \psi_3|),\nonumber\cr
H_{l}&=&(1/\sqrt{2})(\Omega_2|gf\rangle|0\rangle_c-\Omega_1|fg\rangle|0\rangle_c)\langle
 \psi_1|)\nonumber\\&&+(1/2)(\Omega_2|gf\rangle|0\rangle_c+\Omega_1|fg\rangle|0\rangle_c)(\langle
 \psi_2|+\langle \psi_3|)+H.c.,\nonumber\cr
H_{de}&=&\Delta |\psi_1\rangle \langle
\psi_1|+(\Delta/2)(|\psi_2\rangle+|\psi_3\rangle)(\langle
\psi_2|+\langle \psi_3|),
\end{eqnarray}
where $\{|\psi_1\rangle,|\psi_2\rangle,|\psi_3\rangle\}$ are the
eigenvectors of $H_c$:
\begin{eqnarray}\label{}
|\psi_1\rangle&=&(1/\sqrt{2})(|ge\rangle|0\rangle_c-|eg\rangle|0\rangle_c),\nonumber\cr
|\psi_2\rangle&=&(1/2)(|ge\rangle|0\rangle_c-\sqrt{2}|gg\rangle|1\rangle_c
+|eg\rangle|0\rangle_c),\cr
|\psi_3\rangle&=&(1/2)(|ge\rangle|0\rangle_c+\sqrt{2}|gg\rangle|1\rangle_c
+|eg\rangle|0\rangle_c),\nonumber\cr
\end{eqnarray}
and the corresponding eigenvalues are $\lambda_1=0$,
$\lambda_2=-\sqrt{2}g$ and $\lambda_3=\sqrt{2}g$.

Next, we assume that $U_{c} =e^{-iH_c t}$ is the unitary time
evolution operator with respect to the Hamiltonian $H_c$. Assuming
$\Omega_1 = \Omega_2 = \Omega$, after a calculation in the
intermediate "picture", we obtain

\begin{eqnarray}\label{}
H_{l}^{I}&=&U_{c}^{\dag}H_{l}U_{c}=(\Omega/\sqrt{2})(|gf\rangle|0\rangle_c
-|fg\rangle|0\rangle_c)\langle\psi_1|\nonumber\cr
&&+(\Omega/2)(|gf\rangle|0\rangle_c+|fg\rangle|0\rangle_c)(\langle
 \psi_2|e^{i\sqrt{2}g t}+\langle \psi_3|e^{-i\sqrt{2}g t})\nonumber+H.c.,\nonumber\cr\cr
H_{de}^{I}&=&U_{c}^{\dag}H_{de}U_{c}=\Delta |\psi_1\rangle
\langle\psi_1|+(\Delta/2)(|\psi_2\rangle\langle
\psi_2|+|\psi_3\rangle\langle \psi_3|)\nonumber\cr
&&+(\Delta/2)(|\psi_2\rangle\langle \psi_3|e^{-i2\sqrt{2}g
t}+|\psi_3\rangle\langle \psi_2|e^{i2\sqrt{2}gt}).
\end{eqnarray}

On the condition that $\Omega << g$ and $\Delta <<2\sqrt{2}g$, the terms in $H_{l}^{I}$ and $H_{de}^{I}$ with high oscillating frequency $\sqrt{2}g$, $2\sqrt{2}g$ can be safely discarded. Since $|\psi_2\rangle$ and $|\psi_3\rangle$ are decoupled from the initial state $|g f\rangle|0\rangle_c$, the terms describing the interaction
associated with there states can be omitted. So the effective
Hamiltonian of the system can be given
\begin{eqnarray}\label{}
H_{eff}&=&(\Omega/sqrt{2})[(|gf\rangle|0\rangle_c-|fg\rangle|0\rangle_c)\langle
 \psi_1|\nonumber\cr&&+|\psi_1\rangle(\langle 0|\langle f g|-\langle 0|\langle g
 f|)]+\Delta |\psi_1\rangle \langle
 \psi_1|.
\end{eqnarray}

If we consider$(|g\rangle|f\rangle-|f\rangle|g\rangle)/\sqrt{2}$ as a stable level $|G\rangle$ , the above equation can be considered as an effective Hamiltonian of the two-level system interacting with a vacuum cavity field. As a result, the stable ground level $|G\rangle$ is coupled to the excited level $|\psi_1\rangle$ with a coupling constant $\Omega$ and a detuning $\Delta$. On the large detuning condition that $|\Delta|\gg\Omega$, there is no transition between $|G\rangle$ and $|\psi_1\rangle$, the coupling only induces a Stark shift for each level. By adiabatically eliminating the excited state $|\psi_1\rangle$, the final effective Hamiltonian governing the evolution of states $|g\rangle|f\rangle|0\rangle_c$ and $|f\rangle|g\rangle|0\rangle_c$ can be rewritten as

\begin{eqnarray}\label{}
H_{eff}&=&(\Omega^2/(2\Delta))[(|g f\rangle \langle f g|+|f
g\rangle \langle g f|)\nonumber\cr
&&-(|g f\rangle \langle g f|+|f
g\rangle \langle f g|)]|0\rangle\langle 0|.
\end{eqnarray}
In order to validate the feasibility of the above physical model, we
perform a direct numerical simulation the Schr\"{o}dinger equation
with the full Hamiltonian in Eq.(1) and the effective Hamiltonian in
Eq.(6). To satisfy the quantum Zeno dynamics $\Omega << g$ and large
detuning $|\Delta|>>\Omega$, we set the parameters $\Omega=0.05g$
and $|\Delta|=0.5g$. We plot the time-dependent populations of the
basic states $|g f\rangle |0\rangle_c$ (P1) and $|f g\rangle
|0\rangle_c$ (P2) governed by the full Hamiltonian in Eq.(1) (green
lines in Fig.2) and the effective Hamiltonian in Eq.(6) (red lines
in Fig.2). It is shown that the population of the basic states
governed by the effective Hamiltonian exhibits excellent agreement
with that governed by the full Hamiltonian when the conditions are
satisfied. Eventually, the simulation result of the
full Hamiltonian is almost the same as that of the effective
Hamiltonian when $\Omega=0.01g$ and $|\Delta| = 0.2g$. It is appropriate that the
above approximation for the Hamiltonian is reliable as long as
$\Delta/\Omega$ is large enough. While the deviations decrease at the cost of the long
evolution time. The larger the scaled ratio $\Delta/\Omega$ is, the
longer the evolution time is. Considering the decoherence, we choose $\Omega=0.05g$
and $|\Delta|=0.5g$ to satisfy the requirement in the following.

\section{Quantum State Transfer}
\label{}

We note that the above model can be used to realize QST. Two NV centers (1 and 2) are coupled to a microcavity. We assume the NV center $1$ is in an arbitrary unknown state $ \alpha\left| g\right\rangle_1 +\beta \left| f\right\rangle_1 $, where
$|\alpha|^{2}+|\beta|^{2}=1$, while the NV center $2$ is prepared in
the state $\left| g\right\rangle_{2}$ and the cavity mode is
initially in vacuum state $\left| 0\right\rangle _c$. The initial
state of the whole system can be described as $\left| \Psi
(0)\right\rangle =(\alpha \left| g\right\rangle_1 +\beta \left|
f\right\rangle_1 )\left| g\right\rangle _2\left| 0\right\rangle _c
$. For the initial state $\left| g\right\rangle _1\left|
g\right\rangle _2\left| 0\right\rangle _c$, it is easily checked
that the evolution is frozen, since $H\left| g\right\rangle _1\left|
g\right\rangle _2\left| 0\right\rangle _c$ = 0. Governed by the
effective Hamiltonian in Eq.(6), and for an interaction time $t$,
the final state of the system becomes

\begin{eqnarray}\label{}
\left| \Psi (t)\right\rangle &=&\alpha \left| g\right\rangle_1\left|g\right\rangle _2\left| 0\right\rangle _c+(\beta/2)[(1-e^{-i(\Omega^2/\Delta)t})\left| g\right\rangle_1 \left|f\right\rangle _2\left| 0\right\rangle _c\nonumber\cr
&&+(1+e^{-i(\Omega^2/\Delta)t})\left| f\right\rangle_1\left| g\right\rangle _2\left| 0\right\rangle _c].
\end{eqnarray}

By selecting the interaction time $t'$ to satisfy
$(\Omega^2/\Delta)t'=\pi$, one will obtain

\begin{eqnarray}\label{}
\left| \Psi (t')\right\rangle =\alpha \left| g\right\rangle_1\left|
g\right\rangle _2\left| 0\right\rangle _c +\beta \left|
g\right\rangle_1 \left| f\right\rangle _2\left| 0\right\rangle _c.
\end{eqnarray}

The quantum state transfer from NV center $1$ to NV center $2$ has
been realized.

To characterize this QST process, we utilize the fidelity which is
given by

\begin{eqnarray}\label{}
F=\left\langle \Psi _{(t)}|\rho| \Psi _{(t)}\right\rangle ,
\end{eqnarray}
where $\left|\Psi _{(t)}\right\rangle$ is the final state described by the Schr\H{o}dinger equation $i(d\left|\Psi_{(t)}\right\rangle/dt)=H\left|\Psi_{(t)}\right\rangle $. Here $H$ is the full Hamiltonian governed by Eq.(1). The large $\Delta$ can ensure the whole system evolution governed by the effective Hamiltonian in Eq.(6). However, the large detuning prolongs the evolution time which will lead to the worse impacts of decoherence. Fig.3 plots the influences of the fluctuation of the detuning ratio $\Delta/g$ and the Rabi frequency ratio $\Omega/g$ on the fidelity of QST. In this scheme, we choose the median values $\Delta/g=0.5$ and $\Omega/g=0.05$, the corresponding fidelity of QST is larger than 0.998. The fluctuation of the detuning ratio $\Delta/g$ and the Rabi frequency ratio $\Omega/g$ almost does not affect the optimal fidelity of QST, which is always larger than 0.982.

In a realistic experiment, the spontaneous emission of the NV
centers and cavity losses on the QST should be taken into account.
In the following part, we present numerical simulation to show how
the dissipation sources take effects. The master equation of the
whole system can be expressed by
\begin{eqnarray}\label{}
\dot{\rho}&=&-i\left[H,\rho \right] +\frac{\kappa}2 \sum_{j=1}^{2}
(2a_{j}\rho a_{j}^{+}-a_{j}^{+}a_{j}\rho -\rho {a}_{j}^{+}a_{j})\nonumber\cr
&&+\frac{\gamma}2\sum_{j=1}^{2}{\sum_{i=f,g}}(2{\sigma }_{ie}^j\rho
{\sigma }_{ei}^j-{\sigma }_{ei}^j{\sigma }_{ie}^j\rho -\rho {\sigma
}_{ei}^j{\sigma }_{ie}^j),
\end{eqnarray}
where $\kappa$ denotes the effective decay rate of the cavity. For
simplicity, we assume $\gamma _{ef}=\gamma _{eg}=\gamma$, and
$\gamma$ represents the branching ration of the spontaneous decay
from level $\left| e\right\rangle $ to $\left| f\right\rangle $ or
$\left| g\right\rangle $. By solving the master equation
numerically, we obtain the relation of the fidelity versus the
scaled ratio $\gamma/g$ and $\kappa/g$ in Fig.4 in the case of
$\Omega=0.05g$ and $|\Delta| = 0.5g$. Fig.4 shows that the fidelity
of QST will decrease slowly with the increasing of cavity decay and
atomic spontaneous emission. The fidelity of QST is insensitive to
cavity decay and atomic spontaneous emission since it is still about
$91\%$ even for $\gamma/g=0.01$ and $\kappa/g=0.2$. The physical
principle behind this phenomenon is that once the quantum Zeno
condition is satisfied, if the cavity is initially in the vacuum
state, no photons are included in the intermediate state
$|\psi_1\rangle$, i.e., the cavity field has not been excited. So the
cavity decay terms in Eq.(10) have a litter influence on the evolution of
the encoded qubit states. What's more, the further large detuning
condition excludes the excited states of NV centres, so this process
is also robust against NV center's spontaneous emission.

\section{ Conditional phase gate}
\label{}

A new type of quantum conditional phase gate(CPG) has been proposed
by Zheng depends neither on dynamical phase shift nor a solid angle
along a suitable loop \cite{S}. Induced by the adiabatic evolution
along dark eigenstates, such type of non-geometric CPG is robust
against moderate fluctuations of experimental parameters. Recently,
Lacour et al \cite{Lacour} gave a scheme to realize arbitrary state
controlled-unitary gate based on fractional stimulated Raman
adiabatic passage. Schemes have also been proposed for reliable
realization of CPG for two atoms \cite{Ye} or NV centers \cite{Xue}
separately trapped in two distant cavities via adiabatic passage.

Now, we will show how to realize a non-geometric CPG via
off-resonant quantum Zeno dynamics. The quantum information is
encoded in ${\left| f\right\rangle ,\left| i\right\rangle}$ for the
NV center $1$, while in ${\left| g\right\rangle,\left| i\right\rangle
}$ for the NV center $2$. The cavity field is initially in vacuum
state. For the initial state $\left| f\right\rangle _1\left|
i\right\rangle _2\left| 0\right\rangle _c$, $\left| i\right\rangle
_1\left| g\right\rangle _2\left| 0\right\rangle _c$ and $\left|
i\right\rangle _1\left| i\right\rangle _2\left| 0\right\rangle _c$,
it is easily checked that the evolution is frozen. If the initial
state of the system is $\left| f\right\rangle _1\left|
g\right\rangle _2\left| 0\right\rangle _c$, considering the phase of
the classical fields, it will evolve in according to Eq.(6),

\begin{eqnarray}\label{}
  \left| f\right\rangle _1\left| g\right\rangle _2\left|0\right\rangle _c & \rightarrow &\frac{e^{i(\varphi_1-\varphi_2)}}{2}[(1-e^{-i(\Omega^2/\Delta)t})\left|g\right\rangle_1 \left| f\right\rangle _2\left| 0\right\rangle _c\nonumber\cr
  &&+(1+e^{-i(\Omega^2/\Delta)t})\left| f\right\rangle_1\left| g\right\rangle _2\left| 0\right\rangle _c]|.
\end{eqnarray}

In step 1, the phases of the laser pulses ($1$ and $2$) are set to
be equal and the interaction time satisfies
$(\Omega^2/\Delta)t_1=\pi$. In step 2, the phase of the laser $2$
takes a circularly polarized $\pi$ rotation, while the phase of the
laser $1$ remains unchanged. After an interaction time $t_{2}=t_1$,
the state of the system will evolve to

\begin{eqnarray}\label{}
 \left| f\right\rangle _1\left| g\right\rangle _2\left|0\right\rangle _c\rightarrow\left| g\right\rangle_1 \left|f\right\rangle _2\left| 0\right\rangle _c \rightarrow e^{i\pi}\left|f\right\rangle _1\left| g\right\rangle _2\left| 0\right\rangle _c.
\end{eqnarray}

Thus, after a total period $T=2\pi\Delta/\Omega^{2}$, $\left|
f\right\rangle _1\left| g\right\rangle _2\left| 0\right\rangle _c$
returns to the initial state with an additional phase shift $\pi$.
Then we obtain
\begin{eqnarray}\label{}
\begin{array}{cccc}
\left| f g \right\rangle\left| 0\right\rangle &\rightarrow& -\left|
f g \right\rangle \left| 0\right\rangle  ,\cr \left| f i
\right\rangle\left| 0\right\rangle &\rightarrow &  \left| f i
\right\rangle\left| 0\right\rangle , \cr \left| i g
\right\rangle\left| 0\right\rangle &\rightarrow& \left| i g
\right\rangle\left| 0\right\rangle , \cr \left| i i
\right\rangle\left| 0\right\rangle &\rightarrow& \left| i i
\right\rangle\left| 0\right\rangle,
\end{array}
\end{eqnarray}
which corresponds to two-qubit conditional phase gate. Taking
advantage of the quantum Zeno dynamics, the cavity mode keeps in
vacuum state during the whole process, so the decay of the cavity is
largely supressed.

The phase gate is exercised by modifying the phases of the laser
pulses, so the fluctuation of the shift time of pulses influences
the fidelity of the phase gate, as is shown in Fig.5. The result
shows the optimal fidelity of conditional phase gate is almost
unaffected even when the shift time $\delta t/t\leq \pm0.1$. The CPG
is free of the laser field strength and insensitive to the
fluctuation of the shift time, which will reduce the difficulty in
the experiment.

We numerically simulate the full Hamiltonian model in Eq.(1) to show
how the dissipation sources take effects, as is shown in Fig.6. The
fidelity of CPG is insensitive to cavity decay and atomic
spontaneous emission since it is still about $94.5\%$ even for
$\gamma/g=0.01$ and $\kappa/g=0.2$.

\section{Entanglement Dynamics of this System}
\label{}

Sabrina et al \cite{Sabrina} have proposed a strategy to fight
against the deterioration of the entanglement using the quantum Zeno
effect in the resonant case. They also have tested in the
off-resonant regime, protecting the entanglement will become more
efficient than that in the resonant limit \cite{F}. In this paper,
We investigate the entanglement dynamics of two qubits using the
off-resonant quantum Zeno dynamics. If the initial state of the
system is $|\Psi (0)\rangle\ =(\alpha | g f\rangle +\beta | f
g\rangle) | 0\rangle _c $, based on the physical model above, the
system will evolve with respect to the effective Hamiltonian in
Eq.(6). For choosing an interaction time $t$ and
$\lambda=\Omega^2/\Delta$, the final state of the system becomes

\begin{eqnarray}\label{}
 |\Psi(t)\rangle &=&e^{-i(\lambda/2)t}[(\alpha\cos\frac{\lambda}{2} t+i\beta\sin\frac{\lambda}{2} t)|gf\rangle\nonumber\cr
 &&+(i\alpha\sin\frac{\lambda}{2}t+\beta\cos\frac{\lambda}{2}t)|fg\rangle]|0\rangle_c.
\end{eqnarray}

In the standard basis, the reduced density matrix, obtained from the
density operator $|\Psi (t)\rangle \langle \Psi (t)| $ after
tracing over the cavity mode degrees of freedom, takes the form
\begin{eqnarray}\label{}
\rho(t)= \frac{1}{4}\left(
\begin{array}{cccc}
0&0&0&0\cr
0&a&b&0\cr
0&c&d&0\cr
0&0&0&0\cr
\end{array} \right)
\end{eqnarray}

where $a=2(\alpha^2 +\beta^2)+2(\alpha^2 -\beta^2)\cos\lambda t$,
$b=4\alpha\beta+2i(\alpha^2 -\beta^2)\sin\lambda t$,
$c=4\alpha\beta-2i(\alpha^2 -\beta^2)\sin\lambda t$, and
$d=2(\alpha^2 +\beta^2)-2(\alpha^2 -\beta^2)\cos\lambda t$.

We use the concurrence $C(t)$ \cite{Wootters}, ranging from 0 for
separable state to 1 for maximally entangled states, to quantify the
amount of entanglement encoded into the two qubit system. The
explicit analytic expression of $C(t)$ can be obtained from the
reduced density matrix of Eq.(16). The concurrence takes a form
\begin{eqnarray}\label{}
C(t)=\sqrt{(ac^{*}+a^{*}b)(cd^{*}+b^{*}d)}/8
\end{eqnarray}

Assuming here the weight factor ratio of initial state $r=\alpha/\beta$,
$r=1$ is for the maximally entangled states. We can get different
$r$ via changing the Rabi frequencies of classical fields
$\Omega_1/\Omega_2$ in Eq.2. We now look at the entanglement
dynamics versus $r$ and the interaction time $gt$, as is shown in
Fig.7. It is observed that the amplitude of concurrence collapse and
revive versus the interaction time. When $r$ is close to 1, the
amplitude of concurrence is larger than 0.9. Particularly, when
$r=1$, the amplitude of concurrence keeps on 1 and is independent of the
interaction time, which is important in QST. The maximum of $C(t)$ can
reach 1 in a specific time, which is immune to the changing of $r$
and $gt$. While when $r\rightarrow 0$ and $r>>1$, the minimum of $C(t)$
becomes slightly. The above results indicate in our model, the QST
can reach a high fidelity in a special time, and that is independent
of weight factor of initial states, while when $r\rightarrow 1$, the
QST keeps on high fidelity which depends on the initial states. Due
to the quantum Zeno dynamics and the large detuning between NV
centers and cavity field, the interaction is a virtual process and the
system only evolves in a close quantum Zeno subspace spanned by
$\{|g f\rangle|0\rangle_c, |fg\rangle|0\rangle_c\}$.  In all the
processes, the excited state of NV centers and cavity field have not
been populated. Only $|g f\rangle|0\rangle_c$ and
$|fg\rangle|0\rangle_c$ have the population, which has been shown in
Fig.1. From the discussions above, it is worth stressing that the
concurrence in this system crucially depends on the initial state.

To check the influence of the classical field and the detuning, we
numerically simulate the concurrence versus
$\lambda=\Omega^2/\Delta$ and the interaction time $gt$, for
$r=1/3$, as is shown in Fig.8.  When $\lambda\rightarrow 0$, the
$C(t)$ is always small ($<0.4$). In this case, the detuning is
too large, there would be no coupling between the NV centers and the
cavity mode, the concurrence only depends on the initial states of
the system ($r$) and is immune to the interaction time. In the
off-resonant regime, such as $\lambda=0.01g$, the high concurrence
can keep on in a long time, i.e., the oscillating period $T$ is long.
In this case, the detuning meets the requirement of our model, so
the time evolution is only in the subspace composed of the initial
states of the system. In the far off-resonant regime, such as
$\lambda=0.05g$, by contrast, the values of $T$ become short. The
shorter $T$ is, the larger the decay rate of concurrence is. In this
case, the system becomes disentangled more easily. From the
discussions above, we can conclude using the off-resonant quantum
Zeno dynamics to protect the entanglement will become more efficient
than that in the resonant case.

\section{Experimental Feasibility and Conclusion}
\label{}

Now, we discuss the schematic setup and the theoretical model of
proposed scheme may be experimentally realized with quantum optical
devices, such as microsphere cavity \cite{Park} and superconducting
resonator \cite{Kubo}. In order to get identical N-V centers, we
adjust the energy levels of different N-V centers by an external
magnetic field. The NV centers are put near the equator of a
microsphere cavity and interact with the cavity via the evanescent
fields. The coupling constant between the NV centers and the cavity
ranges from hundreds of MHz to several GHz in the experiment
\cite{Barclay}. The $Q$ factor of the microsphere cavity can have a
value exceeding $2\times10^{6}$, leading to a photon leakage rate
$\kappa=\omega/Q\sim 2\pi \times 120$ MHz \cite{Maze}. The
spontaneous decay rate of the NV center is $\gamma\sim 2\pi\times
15$ MHz \cite{Santori}. The coupling between N-V centers and a
microsphere cavity has been realized with the relevant cavity QED
parameters $[g, \gamma, \kappa, \Omega,\Delta ]/2\pi= [1, 0.015,
0.12, 0.05,0.5]$ GHz, with which the corresponding fidelity of the
QST and CPG can reach $97\%$. The decoherence time being longer than
$600\mu s$ at room temperature has been observed for individual NV
centers \cite{Mizuochi}. The QST operation time is about $200ns$
with the parameters above, which is shorter than the decoherence
time of NV center.

In summary, based on off-resonant quantum Zeno dynamics, we have
provided a scheme for implementing quantum state transfer and
conditional phase gate between two NV centers strongly coupled to a
microcavity. By numerical calculation, we have demonstrated that the
present scheme is immune to the excited levels and cavity photons.

\section{acknowledge}
This work is supported from the National Natural Science Foundation of China under Grant No. 11405031, Research Program of Fujian Education Department under Grant No. JA14044.

\begin{figure}
 \centering
  \includegraphics[width=0.5\textwidth]{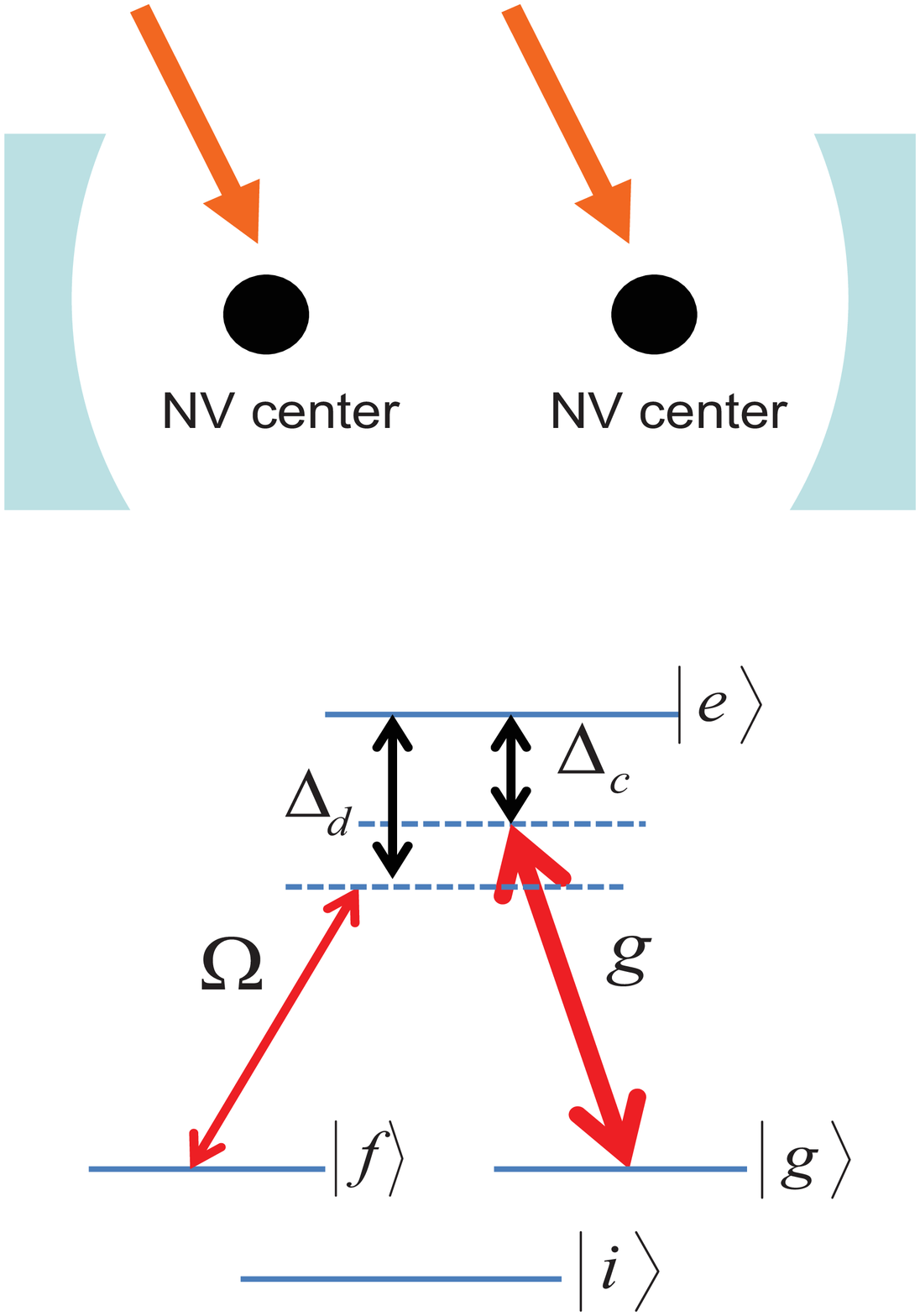}
  \caption{(Color online)The schematic setup for implementing QST and CPG with cavity-NV centers system. Configuration of the NV center level structure and relevant transitions. The ground state $|m_s=+1\rangle=|g\rangle$, $|m_s=-1\rangle=|f\rangle$ and the excited state $\left|A_2\right\rangle=(\left| E_{-}\right\rangle \left|1\right\rangle +\left| E_{+}\right\rangle \left| -1\right\rangle)/{\sqrt{2}}=\left| e\right\rangle$. The state $|m_s=0\rangle=|i\rangle$ is an ancillary state.}
\end{figure}

\begin{figure}
 \centering
  \includegraphics[width=0.5\textwidth]{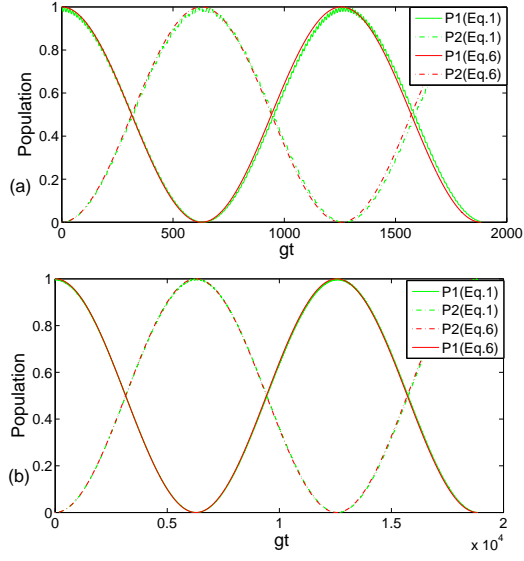}
  \caption{(Color online) The population of the basic states $|g f\rangle |0\rangle_c$ (P1) and $|f g\rangle |0\rangle_c$ (P2) governed by the full Hamiltonian in Eq.(1) (green lines) and the effective Hamiltonian in Eq.(6) (red lines), where (a) $\Omega=0.05g$, $|\Delta| = 0.5g$ (b)$\Omega=0.01g$, $|\Delta| = 0.2g$.}
\end{figure}

\begin{figure}
 \centering
  \includegraphics[width=0.5\textwidth]{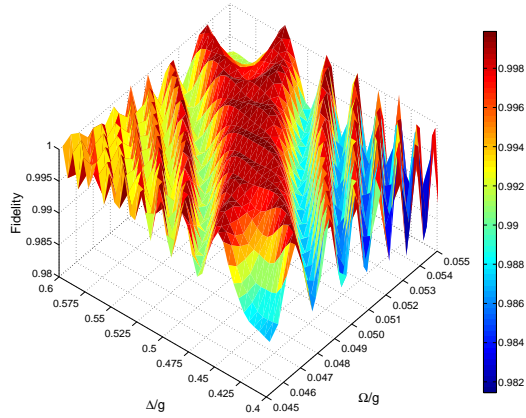}
  \caption{(Color online) The influences of the fluctuation of the detuning ratio $\Delta/g$ and the Rabi frequency ratio $\Omega/g$ on the fidelity of QST, the median values are $\Delta/g=0.5$ and $\Omega/g=0.05$.}
\end{figure}

\begin{figure}
 \centering
  \includegraphics[width=0.5\textwidth]{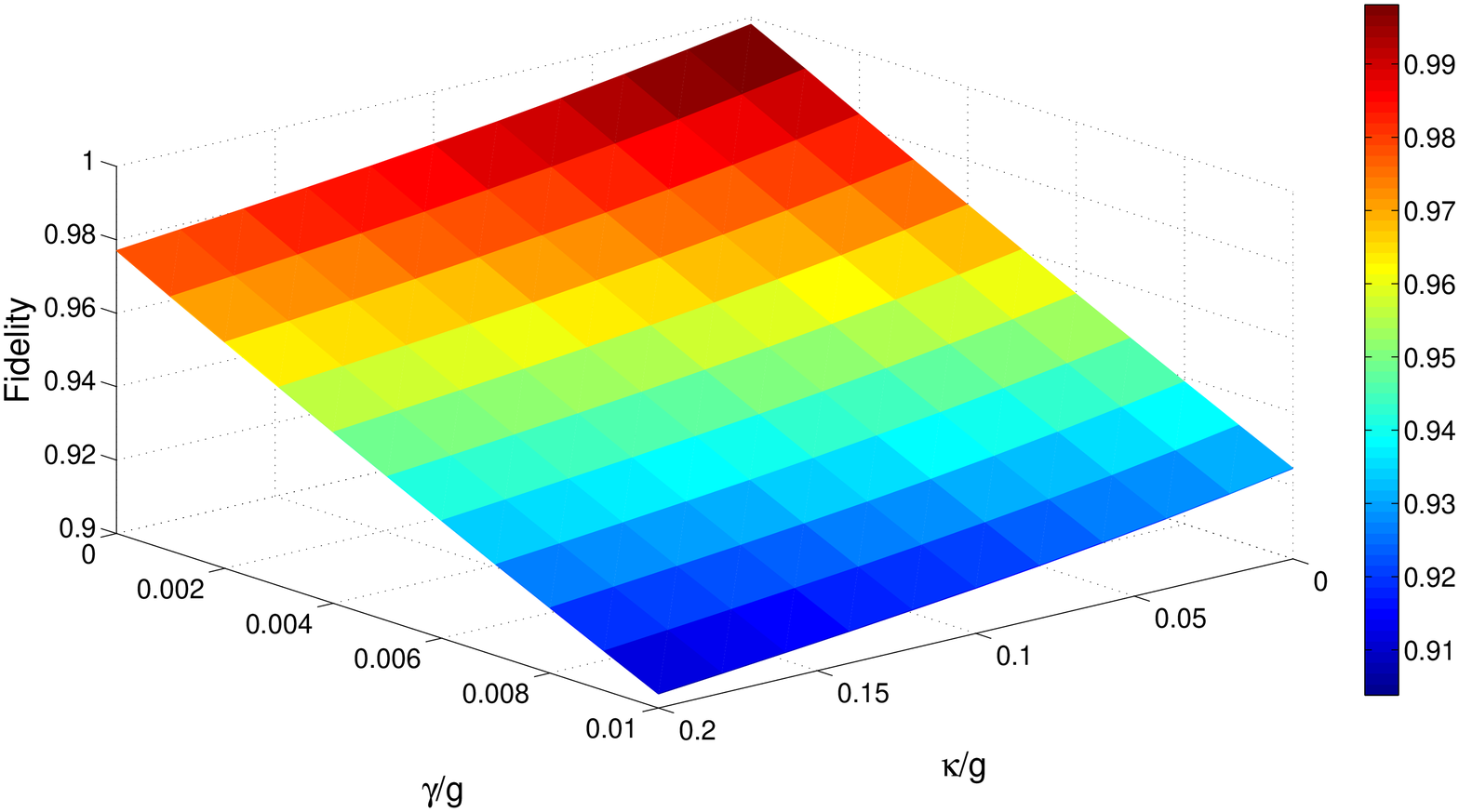}
  \caption{(Color online)The fidelity of QST as a function of the scaled NV center spontaneous emission $\gamma/g$ and scaled cavity decay $\kappa/g$ in the case of $\Omega=0.05g$ and $|\Delta| = 0.5g$. }
\end{figure}

\begin{figure}
 \centering
  \includegraphics[width=0.5\textwidth]{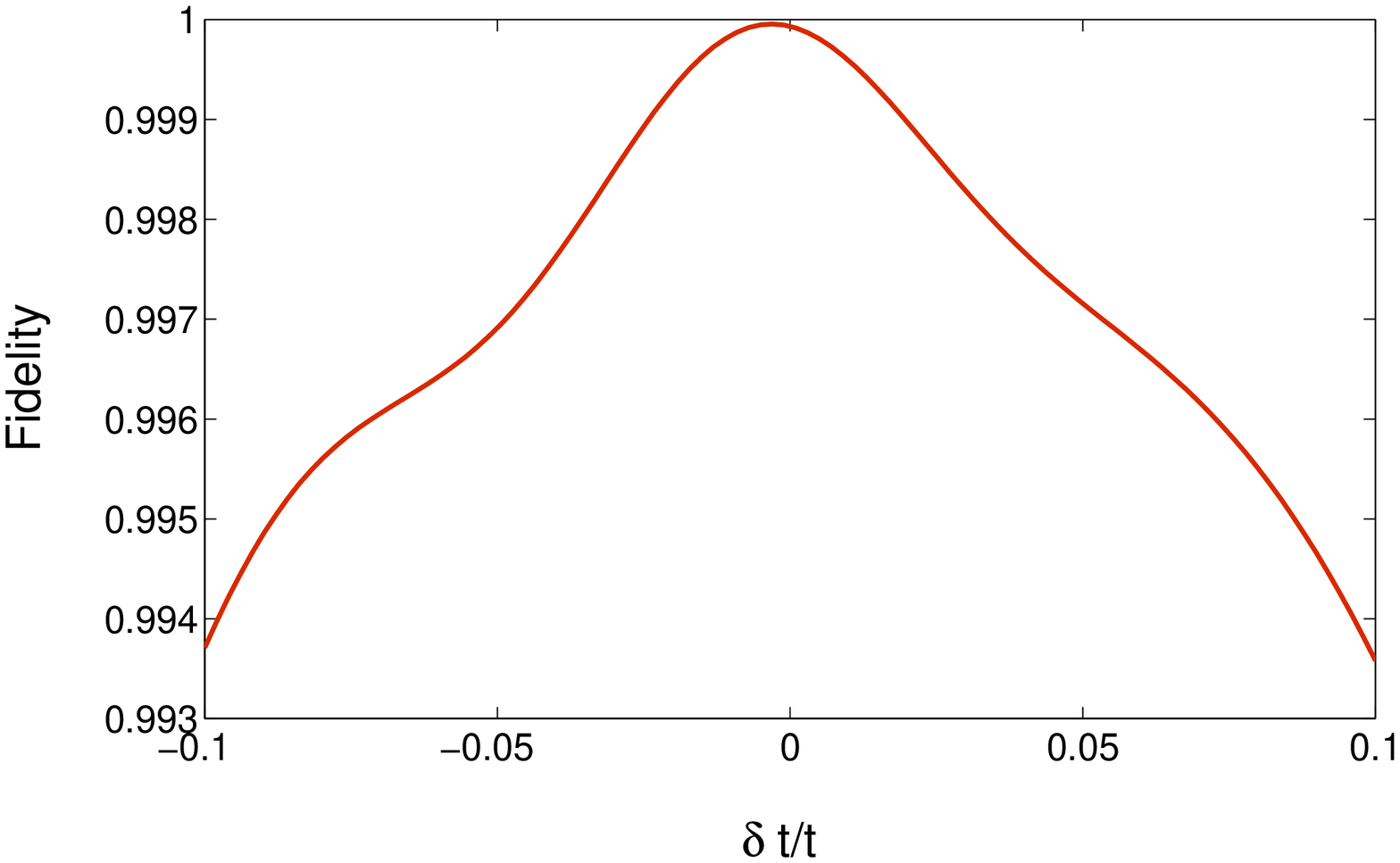}
  \caption{(Color online) The influences of the fluctuation of the shift pulse time $\delta t/t$ on the fidelity of non-geometric CPG, in the case of $\Omega=0.05g$ and $|\Delta| = 0.5g$. }
\end{figure}

\begin{figure}
\centering
  \includegraphics[width=0.5\textwidth]{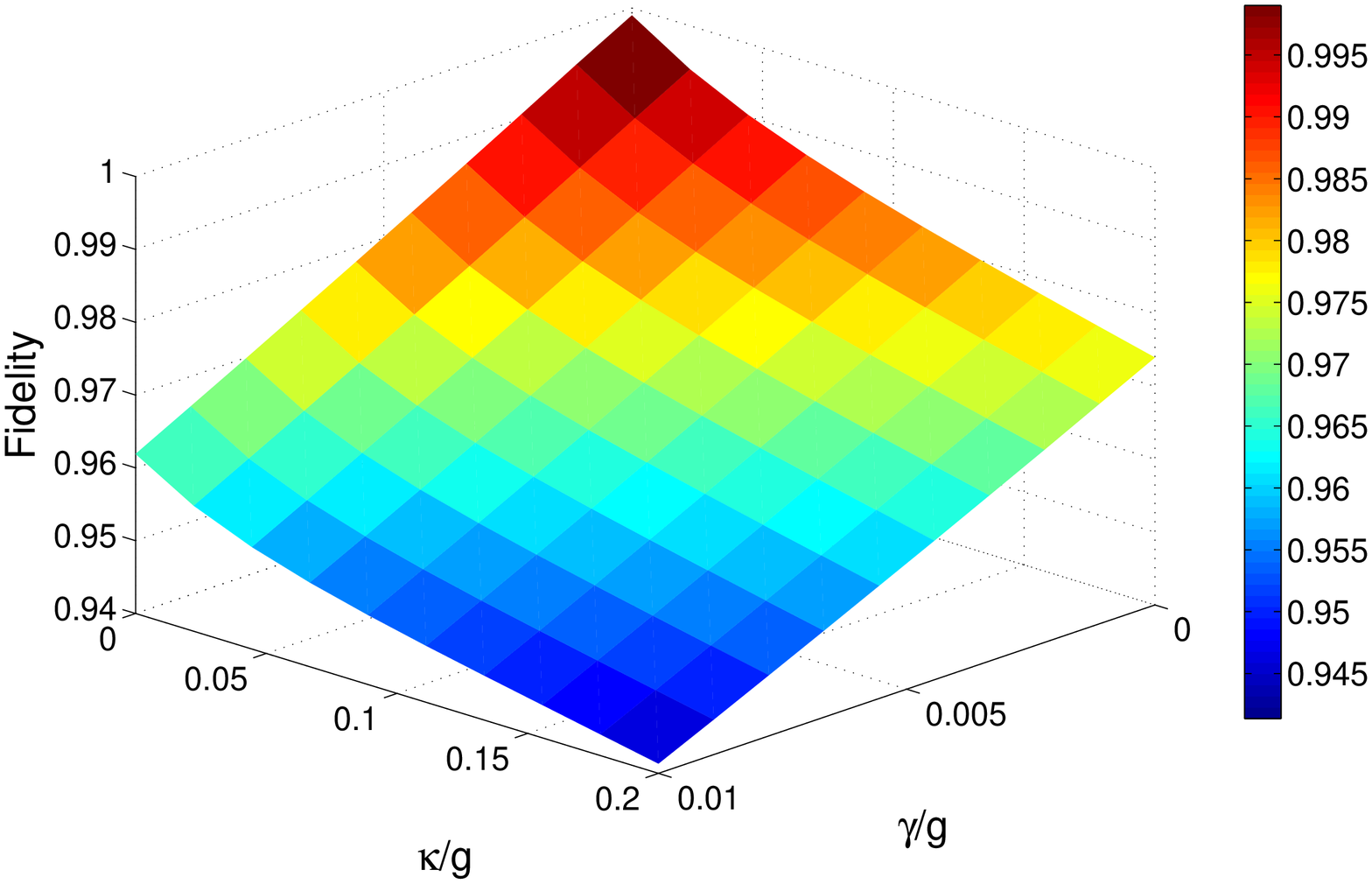}
  \caption{(Color online) The fidelity of non-geometric CPG as a function of the scaled NV center spontaneous emission $\gamma/g$ and scaled cavity decay $\kappa/g$ in the case of $\Omega=0.05g$ and $|\Delta| = 0.5g$. }
\end{figure}

\begin{figure}
 \centering
  \includegraphics[width=0.5\textwidth]{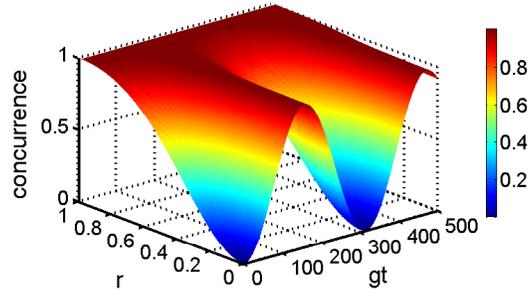}
  \caption{(Color online) The concurrence versus the interaction time $gt$ and the weight factor ratio of initial state $r$. $r=1$ is for the maximally entangled states.}
\end{figure}

\begin{figure}
 \centering
  \includegraphics[width=0.5\textwidth]{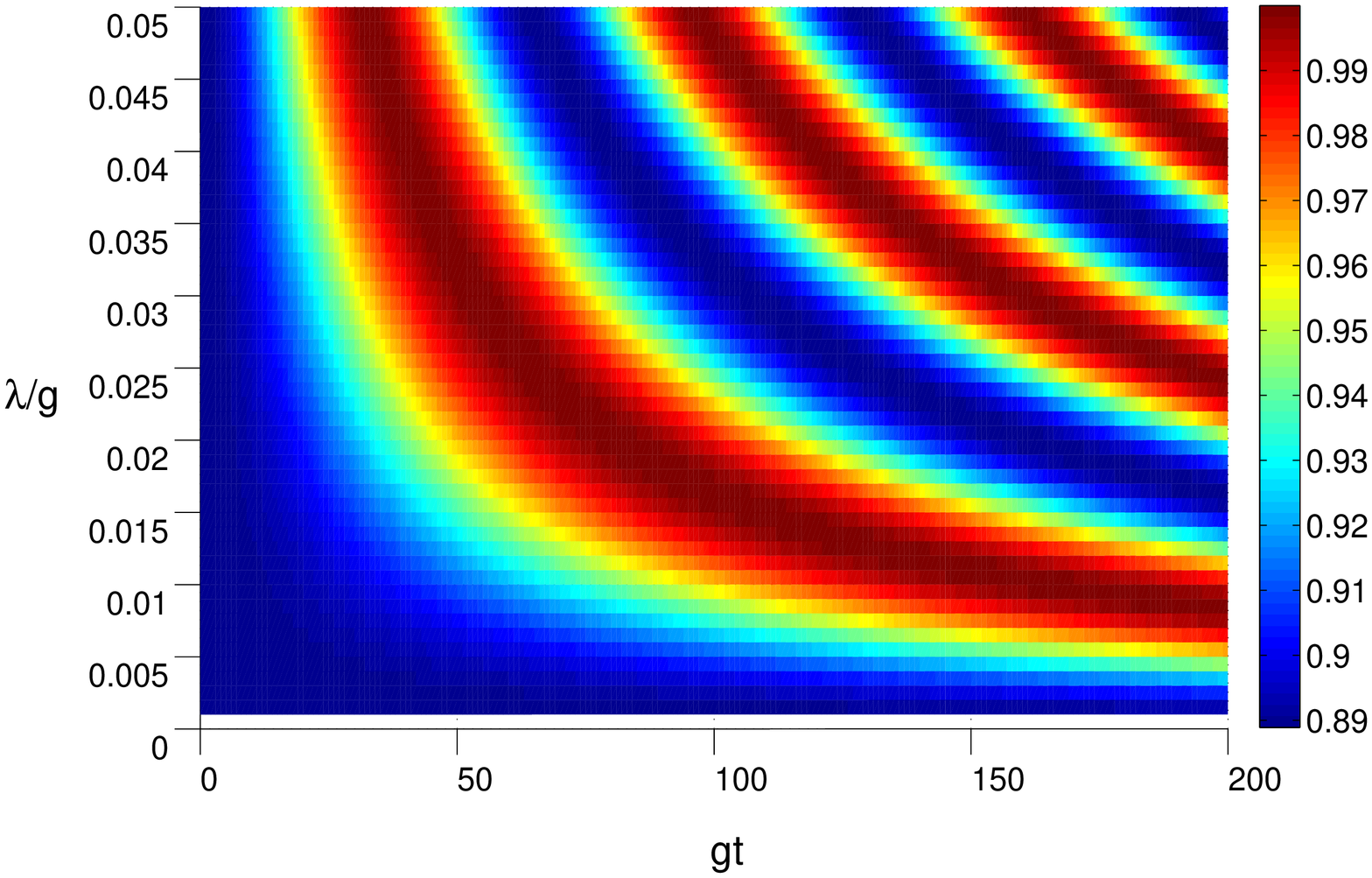}
  \caption{(Color online)The concurrence versus $\lambda/g$ and the interaction time $gt$, for $r=1/3$.}
\end{figure}


\begin{thebibliography}{999}



\bibitem{Lloyd}S. Lloyd, A potentially realizable quantum computer, Science 261 (1993) 1569-1571.

\bibitem{Zheng}S. B. Zheng and G. C. Guo, Efficient scheme for two-Atom entanglement and
 quantum information processing in cavity QED, Phys. Rev. Lett. 85(2000) 2392-2395.

\bibitem{Bose}S. Bose, Quantum Communication through an Unmodulated Spin Chain, Phys. Rev. Lett. 91 (2003) 207901.

\bibitem{Moehring}D. L. Moehring, P. Maunz, S. Olmschenk, et al., Entanglement of single-atom quantum bits at a distance, Nature(London) 449 (2007)68-71.

\bibitem{Kielpinski}D. Kielpinski, C. Monroe and D. J. Wineland, Architecture for a large-scale ion-trap quantum computer, Nature(London) 417 (2002) 709-711.

\bibitem{Blinov}B. B. Blinov, D. L. Moehring, L. M. Duan and C. Monroe,Observation of entanglement between a single trapped atom and a single photon, Nature (London) 428 (2004) 153-157.

\bibitem{Greentree}A. D. Greentree, J. H. Cole, A. R. Hamilton and L. C. L. Hollenberg,Coherent electronic transfer in quantum dot systems using adiabatic passage, Phys. Rev. B 70 (2004) 235317.

\bibitem{yang}C. P. Yang, Quantum information transfer with superconducting flux qubits coupled to a resonator, Phys. Rev. A 82 (2010) 054303.

\bibitem{You}J. Q. You and F. Nori,Quantum information processing with superconducting qubits in a microwave field, Phys. Rev. B 68 (2003) 064509.

\bibitem{Bernien}H. Bernien, B. Hensen, W. Pfaff, G. Koolstra, M. S. Blok, et al., Heralded entanglement between solid-state qubits separated by three metres, Nature 497 (2013)86-90.


\bibitem{Epstein}R. J. Epstein, F. M. Mendoza, Y. K. Kato and D. D. Awschalom, Anisotropic interactions of a single spin and dark-spin spectroscopy in diamond, Nature Phys.1(2005) 94.


\bibitem{Schietinger}S. Schietinger, T. Schroder and O. Benson,One-by-one coupling of single defect centers in nanodiamonds to high-Q Modes of an optical microresonator, Nano Lett. 8 (2008) 3911-3915.


\bibitem{Kimble}H. J. Kimble,The quantum internet, Nature(London) 453 (2008)1023-1030.

\bibitem{Childress}L. Childress, M. V. G. Dutt, J. M. Taylor, et al., Coherent Dynamics of Coupled Electron and Nuclear Spin Qubits in Diamond Science 314 (2006) 281-285.

\bibitem{Sean}S. D. Barrett and K. Pieter, Efficient high-fidelity quantum computation using matter qubits and linear optics, Phys. Rev. A 71(2005) 060310.


\bibitem{Li}P. B. Li, S. Y. Gao and F. L. Li, Quantum-information transfer with nitrogen-vacancy centers coupled to a whispering-gallery microresonator, Phys. Rev. A 83 (2011) 054306.

\bibitem{Ajoy}A. Ashok and C. Paola, Adaptive-boost molecular dynamics simulation of carbon diffusion in iron, Phys. Rev. B 7 (2013) 064303.



\bibitem{Misra}B. Misra and E. C. G. Sudarshan, The Zeno¡¯s paradox in quantum theory, J. Math. Phys.18 (1977) 756.

\bibitem{Cook}R. J. Cook, What are Quantum Jumps, Phys. Scr. T21 (1988) 49-51.

\bibitem{Itano}W. M. Itano, D. J. Heinzen, J. J. Bollinger, et al., Quantum Zeno effect, Phys. Rev. A 41(1990) 2295.

\bibitem{Facchi}P. Facchi and S. Pascazio, Quantum Zeno Subspaces, Phys. Rev. Lett. 89 (2002) 080401.

\bibitem{A}A. S. Zheng, X. Y. Hao and X. Y. L$\ddot{u}$, Generation of three-dimensional entanglement with spin qubits coupled to a bimodal microsphere cavity, J. Phys. B: At Mol. Opt. Phys. 44 (2011) 165507.

\bibitem{R}R. X. Chen and L. T. Shen, Tripartite entanglement of atoms trapped in coupled cavities via quantum Zeno dynamics, Phys. Lett. A 375 (2011) 3840.

\bibitem{Shen}L. T. Shen, H. Z. Wu, Z. B. Yang, Distributed phase-covariant cloning with atomic ensembles via quantum Zeno dynamics, The Eur. Phys. J. D 66 (2012) 123-127.

\bibitem{shi}Z. C. Shi, Y. Xia, J. Song and H. S. Song, Atomic quantum state transferring and swaping via quantum Zeno dynamics, J. Opt. Soc. Am. B 28 (2011) 2909.

\bibitem{F}F. Francica, S. Maniscalco and F. Plastina, Off-resonant quantum Zeno and anti-Zeno effects on entanglement, Phys. Scr. T40(2010) 014044.

\bibitem{shao}S. Zhang, X. Q. Shao, L. Chen, Y. F. Zhao and K. H. Yeon, Robust $\sqrt{swap}$ gate on nitrogen-vacancy centres via quantum Zeno dynamics, J. Phys. B: At. Mol. Opt. Phys. 44 (2011) 075505.

\bibitem{S}S. B. Zheng, Nongeometric conditional phase shift via adiabatic evolution of Dark eigenstates:a new approach to quantum computation, Phys. Rev. Lett. 95(2005)080502.

\bibitem{Lacour}X. Lacour, N. Sangouard, S. Gu¨¦rin1, and H. R. Jauslin, Arbitrary state controlled-unitary gate by adiabatic passage, Phys. Rev. A 73 (2006) 042321.

\bibitem{Ye}S. Y. Ye, S. B. Zheng, Scheme for reliable realization of quantum logic gates for two atoms separately trapped in two distant cavities via optical fibers, Opt. Commun. 281(2008) 1306.

\bibitem{Xue}X. Peng,  A Controlled Phase Gate with Nitrogen-Vacancy Centers in Nanocrystal Coupled to a Silica Microsphere Cavity,Chin. Phys. Lett. 27(2010) 060301.


\bibitem{Sabrina}S. Maniscalco, F. Francica, Rosa L. Zaffino, et al, Protecting Entanglement via the Quantum Zeno Effect, Phys. Rev. Lett. 100(2008) 090503.

\bibitem{Wootters}W. K. Wootters, ntanglement of Formation of an Arbitrary State of Two Qubits, Phys. Rev. Lett. 80 (1998) 2245.

\bibitem{Park}Y. S. Park ,A. K. Cook  and H. L. Wang, Composite optical microcavity of diamond nanopillar and silica microsphere, Nano Lett.9 (2009) 1447-1450.

\bibitem{Kubo}Y. Kubo, C. Grezes, A. Dewes, et al, Hybrid Quantum Circuit with a Superconducting Qubit Coupled to a Spin Ensemble, Phys. Rev. Lett. 107(2011) 220501.


\bibitem{Barclay}P. E. Barclay, C. Santori, K. M. Fu, et al, Coherent interference effects in a nano-assembled diamond NV center cavity-QED system, Opt. Express 19 (2009) 8081.

\bibitem{Maze}J. Maze, P. L. Stanwix, J. S. Hodges, and \emph{et al}., Nanoscale magnetic sensing with an individual electronic spin in diamond , Nature 455 (2008) 644.

\bibitem{Santori}C. Santori, P. Tamarat, P. Neumann, \emph{et al}., Coherent Population Trapping of Single Spins in Diamond under Optical Excitation, Phys. Rev. Lett. 97 (2006) 247401.


\bibitem{Mizuochi}N. Mizuochi, P. Neumann, F. Rempp,\emph{et al.}, Coherence of single spins coupled to a nuclear spin bath of varying density, Phys. Rev. B 80 (2009) 041201.

\end{thebibliography}
\end{document}